\documentclass[reprint,aip,jcp]{revtex4-1}
\usepackage{graphicx}
\usepackage{dcolumn}
\usepackage{bm}
\usepackage{subfigure}
\usepackage{tikz}
\usepackage{color}
\usepackage{lineno}
\usepackage{marginnote}
\usepackage{comment}

\usepackage{wasysym}
\usepackage{amssymb}	

\newcommand{\pd}[2]{\frac{\partial #1}{\partial #2}} 
 
\let\baraccent=\= 
\renewcommand{\=}[1]{\stackrel{#1}{=}} 

\usetikzlibrary{shapes.callouts}
\usetikzlibrary{shapes.geometric,shapes.arrows,decorations.pathmorphing}
\usetikzlibrary{matrix,chains,scopes,positioning,arrows,fit}

\linethickness{0.4mm}
\definecolor{mma1}{rgb}{0.3725,0.5098,0.7020}
\definecolor{mma2}{rgb}{0.8745,0.6078,0.2039}
\definecolor{mma3}{rgb}{0.507813,0.714844,0.2039}
\definecolor{mma4}{rgb}{0.9137,0.3882,0.2398}

\tikzset{
  laser/.style   = { ultra thick, mma4},
  connect/.style = { dashed, red },
  notice/.style  = { draw, rectangle callout, callout relative pointer={#1} },
  label/.style   = { text width=2cm }
}

\begin{document}
\title{Probing dynamical symmetry breaking using quantum-entangled photons}



\author{Hao Li}
\affiliation{Department of Chemistry, University of Houston, Houston, TX 77204}

\author{Andrei Piryatinski}
\affiliation{Theoretical Division, Los Alamos National Lab, Los Alamos, NM 87545}

\author{Jonathan Jerke}
\affiliation{Department of Chemistry and Biochemistry and Department of Physics, Texas Tech University, Lubbock, Texas 79409-1061}
\altaffiliation[Also at ]{Department of Physics, Texas Southern University, Houston, Texas 77004}

\author{Ajay~Ram~Srimath~Kandada}
\affiliation{School of Chemistry \& Biochemistry and School of Physics, Georgia Institute of Technology, 901 Atlantic Drive, Atlanta, Georgia 30332}
\affiliation{Center for Nano Science and Technology @Polimi, Istituto Italiano di Tecnologia, via Giovanni Pascoli 70/3, 20133 Milano, Italy}

\author{Carlos Silva}
\affiliation{School of Chemistry \& Biochemistry and School of Physics, Georgia Institute of Technology, 901 Atlantic Drive, Atlanta, Georgia 30332}

\author{Eric Bittner}
\affiliation{Department of Chemistry \& Department of Physics, University of Houston, Houston, TX 77204}
\email{bittner@uh.edu}

\begin{abstract}

We present an input/output analysis of photon-correlation experiments whereby a quantum
mechanically entangled bi-photon state interacts with 
a material sample placed in one arm of a  Hong-Ou-Mandel (HOM)
apparatus. We show that the output signal contains 
detailed information about subsequent entanglement with
the microscopic quantum states in the sample.
In particular, we apply the method to an ensemble 
of emitters interacting with a common photon mode within 
the open-system Dicke Model. Our results indicate 
considerable dynamical information concerning spontaneous
symmetry breaking can be revealed with such an experimental 
system.

\end{abstract}

\date{\today}%
\maketitle 


\section{Introduction}

The interaction between light and matter lies at the heart of all photophysics and 
spectroscopy.  Typically, one treats
 the interaction within a semi-classical approximation, 
treating light as an oscillating classical electro-magnetic wave as given by Maxwell's equations. 
It is well recognized that light has a quantum mechanical discreteness (photons)
and one can prepare entangled interacting photon states. The pioneering work 
by Hanbury Brown and Twiss in the 1950's, who measured intensity correlations 
in light originating from thermal sources, set the stage for what has become quantum optics.
\cite{BRANNEN1956,
HANBURYBROWN1956,
PURCELL1956,Brown1957,Brown1958,Fano1961,Paul1982}
Quantum photons play a central role in a number of advanced technologies including
quantum cryptography\cite{RevModPhys.74.145},
quantum communications\cite{Gisin:2007aa},
and quantum computation\cite{Knill2001,PhysRevA.79.033832}.
%
Only recently has it been proposed that  entangled photons
can be exploited as  a useful spectroscopic probe of atomic and molecular processes.
\cite{PhysRevA.79.033832,Lemos2014,Carreno2016,Carreno2016a,Kalashnikov2016a,Kalashnikov2016b}

The spectral and temporal nature of entangled photons offer a unique means for interrogating
the dynamics and interactions between molecular states.
The crucial consideration is that when entangled photons are created, typically by spontaneous parametric down-conversion, 
there is a precise relation between the frequency and wavevectors of the entangled pair.  
For example if we create two entangled photons from a common 
laser source, energy conservation dictates that $\omega_{laser} = \omega_{1} + \omega_{2}$.  Hence 
measuring the frequency  of either photon will collapse the quantum entanglement and the
frequency of the other photon will be precisely defined.   Moreover, in the case 
of multi-photon absorption, entangled 2-photon absorption is greatly enhanced relative to 
classical 2-photon absorption since the cross-section scales linearly rather than quadratically with intensity. 
Recent work by Schlawin {\em et al.}  indicate that entangled photon pairs may be useful in controlling and 
manipulating population on the 2-exciton manifold of a  model biological energy transport
 system. \cite{Schlawin2013}
  The non-classical features of entangled photons have also been used as a highly sensitive 
detector of ultra-fast emission from organic materials.  
\cite{doi:10.1021/acs.jpclett.6b02378}
Beyond the potential practical applications of quantum light in 
high-fidelity communication and quantum encryption, 
by probing systems undergoing 
spontaneous symmetry breaking with quantum photons one can 
draw analogies between bench-top laboratory based experiments and 
experimentally inaccessible systems such as black holes, the early 
Universe, and cosmological strings.\cite{Zurek1985,Nation:2010}

\begin{figure}[b]
\includegraphics[width=\columnwidth]{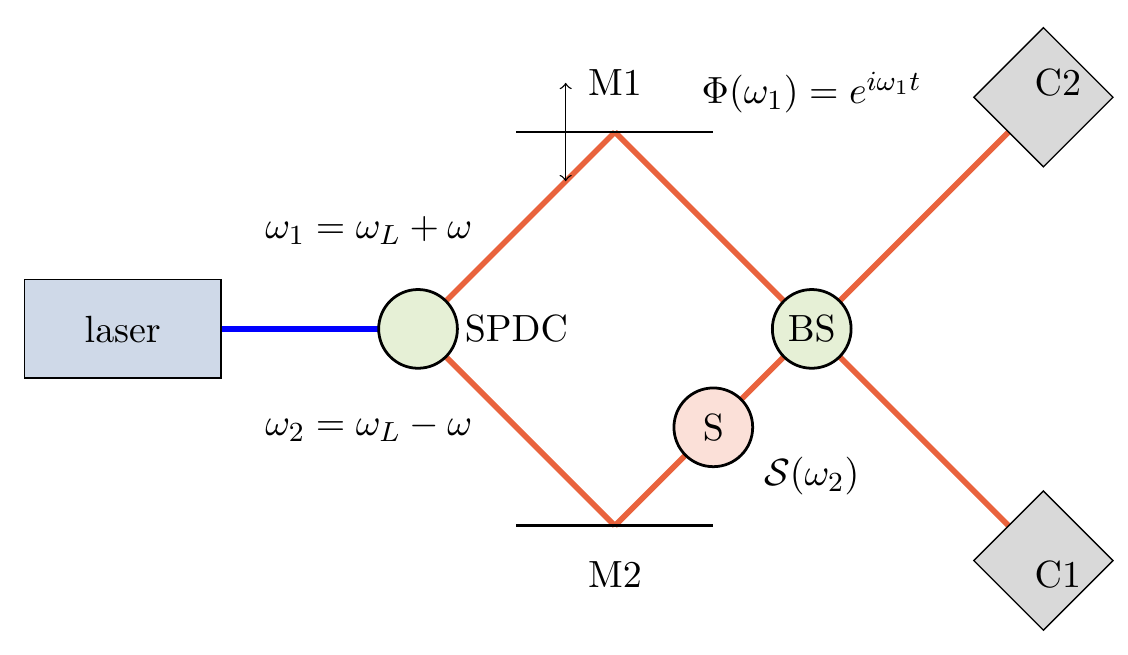}
\caption{{\bf Sketch of Hong-Ou-Mandel apparatus (HOM) for 2-photon coincidence detection.  }
The initial laser beam pass through a spontaneous parametric down-conversion crystal
(SPDC) creating an entangled photon pair which is 
split  into idler ($\omega_1$) and signal ($\omega_2$) modes. 
The two beams are  subsequently reflected back towards a 
beam-splitter (BS) by mirrors M1 and M2 and the signal mode further interacts with 
a sample at S. The modes are recombined by a beam-splitter (BS) and 
directed towards coincidence counters C1 and C2. 
Not shown in our sketch is an optional pumping laser for creating 
a steady-state exciton density in S.
}\label{wwa}
\end{figure}

In this paper, we  provide a precise connection between a material sample, described in terms of a model Hamiltonian,  and the  resulting signal
in the context of the interference  experiment described by Kalashnikov {\em et al.} 
in Ref.\cite{Kalashnikov2016b}, using the Dicke model for an ensemble
of two-level atoms as input.\cite{PhysRev.93.99} The Dicke model
is an important test-case since it undergoes a dynamical phase transition.
Such behaviour was recently observed by Klinder {\em et al.} by coupling 
an atomic Bose-Einstein condensate to an optical cavity.\cite{kilnder:2015}
We begin with a brief overview of the 
photon coincidence experiment and the preparation of two-photon 
entangled states, termed ``Bell-states''.  We then use the input/output approach of Gardner and Collett
\cite{Gardiner1985}  to develop a
means for computing the transmission function for a 
–material system placed in one of the arms of the Hong-Ou-Mandel (HOM) apparatus sketched in Fig.~\ref{wwa}.



\section{Quantum interference of entangled photons}

We consider the interferometric scheme implemented by Kalashnikov{\em et al.} \cite{Kalashnikov2016b}
A CW laser beam is incident on a nonlinear crystal, creating an entangled photon pair
state by spontaneous parametric down-conversion (SPDC), which we shall denote as a Bell state
\begin{eqnarray}
|\bell_{1} \rangle &=& \iint d\omega_{1} d\omega_{2} {\cal F}(\omega_{1},\omega_{2})
B^{\dagger}_{S}(\omega_{1})
B^{\dagger}_{I}(\omega_{2})
|0\rangle,
\end{eqnarray}
where ${\cal F}(\omega_{1},\omega_{2})$ is the bi-photon field amplitude and 
$B^{\dagger}_{S,I}(\omega_{i})$ creates a photon with frequency $\omega_{i}$ in either the signal or idler branch. 
The ket $|0\rangle$ is the vacuum state and $|\omega_{1}\omega_{2}\rangle$ 
denotes a two photon state.

In general, energy conservation requires that the entangled photons 
generated by  spontaneous parametric down conversion (SPDC) obey
 $\omega_{L} = \omega_{1} + \omega_{2}$.  Similarly, conservation of photon 
momentum requires  ${\bf k}_{L} = {\bf k}_{1} + {\bf k}_{2}$.   By manipulating
the SPDC crystal, one can generate entangled photon pairs with 
different frequencies.  
As a result, the bi-photon field is strongly anti-correlated in frequency with 
\begin{eqnarray}
|\bell\rangle = \int dz {\cal F}(z)B_{1}^{\dagger}(\omega_{L}-z)B_{2}^{\dagger}(\omega_{L}+z)|0\rangle,
\end{eqnarray}
where $\omega_{L}$ is the central frequency of the bi-photon field.
This aspect was recently exploited in Ref. \cite{Kalashnikov2016a},
which used a visible photon in the idler 
branch and an infrared (IR) 
photon in the signal branch, interacting with the sample. 

As sketched in Figure~\ref{wwa}, both signal and idler are 
reflected back towards a beam-splitter
(BS) by mirrors M1 and M2.  
M1 introduces an optical delay with transmission function $\Phi(\omega)$ 
which we will take to be of modulo 1.   In the other arm, 
we introduce a resonant medium at S with transmission function ${\cal S}(\omega)$. 
Not shown in our sketch is an optional pumping laser for creating 
a steady state exciton density in S.

%

Upon interacting with both the delay element and the medium, the Bell-state
can be rewritten as
\begin{eqnarray}
 |\bell_{2} \rangle = \iint d\omega_{1} d\omega_{2} F(\omega_{1},\omega_{2})
B^{\dagger}_{I}(\omega_{1})
B^{\dagger}_{S}(\omega_{2})
\Phi(\omega_{1}){\cal S}(\omega_{2})
|0\rangle.
\nonumber 
\\
\end{eqnarray}
Finally, the two beams are recombined by a beam splitter (BS) and the 
coincidence rate is given by 
\begin{eqnarray}
P_{c} &=& \iint d\omega_{1}  d\omega_{2} |\langle\omega_{1}\omega_{2} | \psi_{c} \rangle | ^{2} \nonumber
\\
&=&
\frac{1}{4}\iint d\omega_{1}  d\omega_{2}
\left\{
|{\cal F}(\omega_{1},\omega_{2}) {\cal S}(\omega_{2})|^{2} +
|{\cal F}(\omega_{2},\omega_{1}) {\cal S}(\omega_{1})|^{2}
\right.
\nonumber \\
&-&
\left.
2 {\rm Re}\left[{\cal F}^{*}(\omega_{1},\omega_{2}){\cal F}(\omega_{2},\omega_{\textcolor{black}{1}})
{\cal S}^{*}(\omega_{2}){\cal S}(\omega_{1})
\Phi^{*}(\omega_{1})\textcolor{black}{\Phi}(\omega_{2})
\right]
\right\}.
\label{eq:pc}
\end{eqnarray}

We assume that the delay stage is dispersionless with $\Phi(\omega) = e^{i\omega t}$ 
and re-write equation~\ref{eq:pc}  as 
\begin{eqnarray}
P_{c}(t_{\rm {delay}})
&=&
\frac{1}{4}\int_{-\infty}^{+\infty} dz
|{\cal F}(z)|^{2}
\left\{
|{\cal S}(\omega_{L}-z)|^{2} +|{\cal S}(\omega_{L}+z)|^{2}
\right.
\nonumber 
\\
&-& \left. 2 {\rm Re}\left[
{\cal S}^{*}(\omega_{L}-z)
{\cal S}(\omega_{L}+z)
e^{-2iz t_{\rm delay}}
\right]
\right\},
\label{eq:pc2}
\end{eqnarray}
where $t_{\rm delay}$ is the time lag between entangled
 photons traversing  the upper and lower arms of the HOM apparatus. 
This is proportional to the counting rate of coincident photons 
observed at detectors C1 and C2 and serves as the 
central experimental observable.


%
%
\begin{figure*}[t]
\includegraphics[width=\columnwidth]{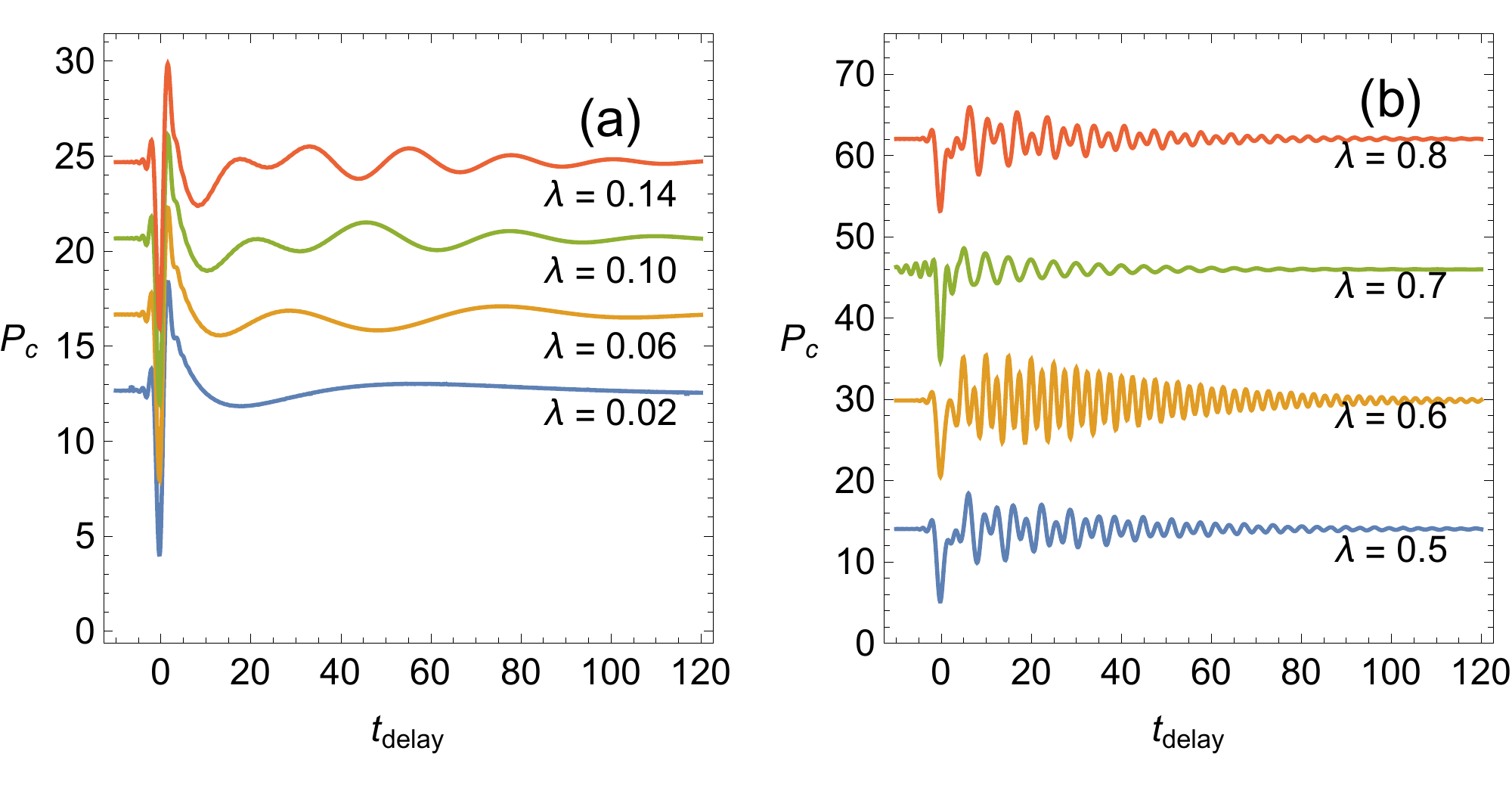}
\caption{
{\bf Photon Coincidence Rates vs. Coupling.}
We compare here the computed (relative) coincidence counting rates 
as $\lambda$ increases from weak (a) to strong coupling (b).  As throughout this 
work we take $\lambda_{c} = 0.7516$ as per equation~\ref{lambda-crit}.  Each scan
is shifted vertically for clarity.}
\label{pc-scan-lambda}
\end{figure*}

\section{Results}
A crucial component of our approach is the action of the sample at S which introduces a 
transmission function $S(\omega)$ into the final Bell state.   We wish to connect this 
function to the dynamics and molecular interactions within the sample.  To accomplish this, we 
use the input/output formulation of quantum optics and apply this to 
an ensemble of identical 2-level states coupled to a common photon mode.\cite{PhysRevA.75.013804}
Technical details of our approach are presented in the Methods section of this paper.
In short, we begin with a description of the material system described by $N$  two-level spin states coupled to 
common set of photon cavity modes. 
 \begin{eqnarray}
\hat H_{sys} &=&\sum_{j} \frac{\hbar\omega_{o}}{2}\hat\sigma_{z,j}
+ \sum_{k}\hbar(\omega_{k}-i\kappa)\hat\psi_{k}^{\dagger}\hat\psi_{k}  \nonumber \\
&+&\sum_{k,j}\frac{\hbar\lambda_{kj}}{\sqrt{N}}(\hat\psi_{k}^{\dagger} + \hat\psi_{k})(\hat\sigma^{+}_{j} + \hat\sigma_j^{-})\label{dickeH-1}
\end{eqnarray}
where $\{\hat\sigma_{z,j},\hat\sigma^{\pm}_{j}\}$ are local spin-1/2 operators for site $j$, $\hbar\omega_{o}$ is the local 
excitation energy, and $\lambda_{kj}$ is the coupling between the $k$th photon mode and the $j$th site, which we will take to be
uniform over all sites.   We introduce $\kappa$ as the decay rate of a cavity photon.
We allow the photons in the cavity (S)  to exchange quanta with photons in the HOM apparatus 
and derive the Heisenberg equations of motion corresponding to input and output photon fields within a 
steady state assumption. This allows us to compute the scattering matrix connecting an 
incoming photon with frequency $\nu$ from the field to an outgoing photon with frequency $\nu$ returned to the 
field viz. 
\begin{eqnarray}
\Psi_{out}(\nu) = -\hat\Omega^{(-)\dagger}_{out}\hat\Omega^{(+)}_{in}\Psi_{in}(\nu)
\end{eqnarray}
where $\hat\Omega^{(\pm)}_{in,out}$ are M{\o}ller operators that propagate an incoming (or outgoing) state from $t\to-\infty$ to $t=0$ 
where it interacts with the sample. 
or from $t=0$ to an outgoing (or incoming) state at $t\to+\infty$ and give the $S$-matrix in the form of a response function
\begin{eqnarray}
{\cal S}(\nu) = \langle \delta \Psi^{\dagger}_{out}(\nu)\delta \Psi_{out}(\nu')\rangle\delta(\nu-\nu')
\label{response-main}
\end{eqnarray}
where the $\delta \Psi_{out}(\nu)$ are fluctuations in the output photon field about a steady-state solution. 
The derivation of $S(\nu)$ for the Dicke model and its incorporation into equation~\ref{eq:pc2}
is a central result of this work and is presented in the Methods section of this paper. 
In general, $S(\nu)$ is a complex function with a series of poles displaced above the 
real $\nu$ axis and we employ a sync-transformation method to integrate equation~\ref{eq:pc2}.
The approach can be applied to any model Hamiltonian system and provides the necessary connection between a microscopic model and its predicted photon coincidence.

Before discussing the results of our calculations, it is important to recapitulate 
a number of aspects of the Dicke model and how these features are manifest in the 
photon coincidence counting rates.   As stated already, 
we assume that the sample is in a steady state by 
exchanging the photons in the HOM apparatus  
with photons within the sample cavity and that ${\cal S}(\nu)$ can be
described within a linear-response theory.   
Because the cavity photons become entangled with 
the material excitations, the excitation frequencies are split into upper photonic ($\omega_{+}$) 
and lower excitonic ($\omega_{-}$)  branches. 
These frequencies are complex corresponding to 
the exchange rate between cavity and HOM photons. 

At very low values of $\lambda_{k}$, the real frequencies  are equal 
and $\omega_{+} = \omega_{-}$.  In this over-damped regime, 
photons leak from the cavity before the photon/exciton state has undergone a single Rabi oscillation.  
At $\lambda_{k} = \kappa/2$ the system becomes critically damped and for $\lambda_{k}<\kappa/2$
and the degeneracy between the upper and lower polariton branches is lifted. 
As $\lambda_{k}$ increases above a critical value given by 
\begin{eqnarray}
\lambda_{c} = \sqrt{\frac{\omega_{k}\omega_o}{4}\left(1 + \frac{\kappa^{2}}{\omega_{k}^{2}}\right)},
\label{lambda-crit}
\end{eqnarray}
the system undergoes a quantum phase transition  when $\omega_{-} = 0$.
Above this regime, excitations from the 
non-equilibrium steady-state become collective and super-radiant. 
For our numerical results, unless otherwise noted we use dimensionless quantities,
 taking $\omega_{o} = \omega_{k} = 1.5$ for both the exciton frequency and cavity mode frequency,
  $\kappa = 0.05$ for the cavity decay. These give 
a critical value of $\lambda_{c} = 0.7516$.


We first consider the  photon coincidence in the normal regime. 
 Figures~\ref{pc-scan-lambda}(a,b) show the variation of the photon coincidence count when the 
laser frequency is resonant with the excitons ($\omega_{k} = \omega_{o}$).  
For low values of $\lambda_{k}$, the system is in the over-damped regime and
the resulting coincidence scan reveals a slow decay for positive values of the 
time-delay.   This is the perturbative regime in which the scattering photon is 
dephased by the interaction with the sample, but there is insufficient time 
for the photon to become entangled with the sample.  
For $\lambda_{k} > \kappa/2$, the scattering photon is increasingly 
entangled with the material and further oscillatory structure begins to emerge
in the coincidence scan.   

\begin{figure*}[t]
\includegraphics[width=\columnwidth]{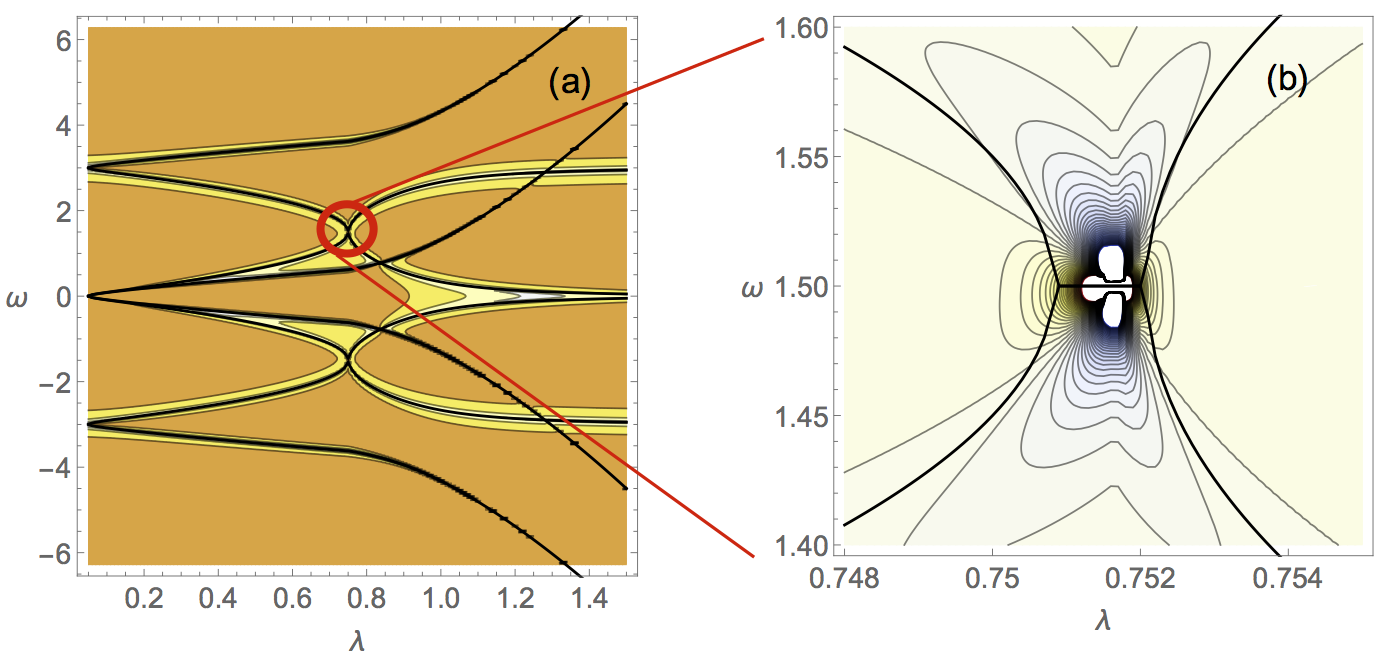}
\caption{{\bf Pole Structure of Response Function}.
(a) Contour plot of ${\cal P}_{c}(\omega)$~(Eq.\ref{scatamp})
 varying the coupling $\lambda$.  Superimposed are the locations of the (real) poles of the spectral response ${\cal S}^*(\omega_L-\omega){\cal S}(\omega_L+\omega)$.
(b) Expansion of the intersection between $\omega_{-}$ branches in the critical regime showing the rapid variation of the 
response in this region.  
}\label{poles}
\end{figure*}

In the strong coupling regime, $P_{c}(t)$ becomes increasingly oscillatory with contributions from
multiple frequency components.
The origin of the structure is further revealed upon taking the Fourier cosine transform of
$P_{c}(t)$ (equation~\ref{eq:pc2})  taking the 
bandwidth of the bi-photon amplitude to be broad enough to span the full spectral range.
The first two terms in the integral of equation~\ref{eq:pc2} are independent of time and simply give a
background count and can be ignored for purpose of analysis. 
The third term depends upon the time delay and is Fourier-cosine transform of 
the bi-photon amplitude times the scattering amplitudes,
\begin{eqnarray}
{\cal P}_{c}(\omega) = |{\cal F}(\omega)|^{2}{\cal S}^{*}(\omega_{L}-\omega){\cal S}(\omega_{L}+\omega). \label{scatamp}
\end{eqnarray}
As we show in the Methods, ${\cal S}(\omega)$ has a series of poles on the complex
plane that correspond to the 
frequency spectrum of  fluctuations
 about the matter-radiation steady-state as given by the eigenvalues of ${\cal M}_{s}$ in 
 equation~\ref{modelM} in the Methods.
 In Figure~\ref{poles}(a,b) we show the evolution of pole-structure of  
${\cal S}^{*}(\omega_{L}-z){\cal S}(\omega_{L}+z)$ 
superimposed over the Fourier-cosine transform of the coincidence counts (${\cal P}_{c}(\omega)$)
 with increasing coupling $\lambda_{k}$ revealing that
 that both excitonic and photonic branches contribute to the overall photon coincidence
 counting rates.  

A close examination of the pole structure in the vicinity of the 
phase transition reveals that two of the 
$\omega_k^{(-)}$ modes become degenerate 
over a small range of $\lambda_{k}$ but with
different imaginary components 
indicating that the two modes decay at different rates. 
This is manifest in Figure~\ref{poles}b by the rapid variation
and divergence in the ${\cal S}^{*}(\omega_{L}-z){\cal S}(\omega_{L}+z)$ about $\lambda_c$. \cite{Kopylov2013}  While the
 parametric width of this regime is small, it depends entirely
upon the rate of photon exchange between the cavity and the
laser field ($\kappa$).

\section{Discussion}

We present here a formalism and method for connecting
the photon coincidence signals for a sample placed
in a HOM apparatus to the optical response 
of the coupled photon/material system. 
Our formalism reveals that by taking the 
Fourier transform of the $P_{c}(t)$ coincidence 
signal reveals the underlying pole
 structure of the entangled material/photon system.  
Our  idea hinges upon an assumption that the 
interaction with the material preserves the 
initial  entanglement between the two photons and that
sample on the entanglement introduces an additional phase
lag to one of the photons which we formally introduce in the 
form of a scattering response function ${\cal S}$.
The pole-structure in the output comes about from the 
further quantum entanglement of the signal photon with the
sample.  

Encoded in the time-delay signals is important  
information concerning the inner-workings of a 
quantum phase transition.  
Hence, we conclude that entangled photons with 
interferometric detection techniques provide 
a viable and tractable means to extract precise 
information concerning light-matter interactions. 
In particular, the approach reveals that at the onset
of the symmetry-breaking transition between normal and super-radiant
phases, two of the eigenmodes of the light-matter state 
exhibit distinctly different lifetimes.  This signature of
an intrinsic aspect of light-matter entanglement may be
observed in a relatively simple experimental geometry with 
what amounts to a {\em linear} light-scattering/interferometry 
set up.

At first glance, it would appear that using quantum photons 
would not offer a clear advantage over more standard  
spectroscopies based upon a semi-classical description of 
the radiation field.  However, the entanglement variable
adds an additional dimension to the experiment allowing one to 
preform what would ordinarily be a non-linear experiment using classical light
as a linear experiment using quantized light.  
The recent works by Kalashnikov {\em et al.} that inspired this work 
are perhaps the proverbial tip of the iceberg. 
\cite{Kalashnikov2016a,Kalashnikov2016b}




\section*{Acknowledgments}
The work at the University of Houston was funded in
part by the  National Science Foundation (CHE-1664971, MRI-1531814)
and the Robert A. Welch Foundation (E-1337). 
AP acknowledges the support provided by Los Alamos National Laboratory Directed Research and Development (LDRD) Funds.
CS acknowledges support from the School of Chemistry \& Biochemistry and the College of Science of Georgia Tech. 
JJ acknowledges the support of the Army Research Office (W911NF-13-1-0162).
ARSK acknowledges funding from  EU Horizon 2020 via Marie Sklodowska Curie Fellowship (Global) (Project No. 705874).





\appendix

\section{Preparation of Entangled States}
We review here the preparation of the entangled states as propagated to the 
coincidence detectors in Fig. 1. 
The initial laser beam 
produces an entangled photon pair by spontaneous parametric down-conversion at B, with 
state which we shall denote as a Bell state
\begin{eqnarray}
|\bell_{1} \rangle &=& \iint d\omega_{1} d\omega_{2} {\cal F}(\omega_{1},\omega_{2})
B^{\dagger}_{I}(\omega_{1})
B^{\dagger}_{S}(\omega_{2})
|0\rangle \\
&=& \iint d\omega_{1} d\omega_{2} {\cal F}(\omega_{1},\omega_{2})|\omega_{1}\omega_{2}\rangle,
\end{eqnarray}
where ${\cal F}(\omega_{1},\omega_{2})$ is the bi-photon field amplitude and 
$B^{\dagger}_{I,S}(\omega_{i})$ creates a photon with frequency $\omega_{i}$ in either the idler (I) or signal (S) arm of the HOM 
apparatus. 
The ket $|0\rangle$ is the vacuum state and $|\omega_{1}\omega_{2}\rangle$ 
denotes a two photon state.
The two photons are  reflected back towards a 
beam-splitter (BS) by mirrors M1 and M2.  
M1 introduces an optical delay with transmission function $\Phi(\omega)$ 
which we will take to be of modulo 1.   In the other arm, 
we introduce a resonant medium at S with transmission function ${\cal S}(\omega)$. 
%
%
\begin{widetext}
Upon interacting with both the delay element and the medium, the Bell-state
can be rewritten as
\begin{eqnarray}
 |\bell_{2} \rangle = \iint d\omega_{1} d\omega_{2} {\cal F}(\omega_{1},\omega_{2})
B^{\dagger}_{I}(\omega_{1})
B^{\dagger}_{S}(\omega_{2})
\Phi(\omega_{1})S(\omega_{2})
|0\rangle.
\nonumber 
\\
\end{eqnarray}
Finally, the two beams are re-joined by a beam-splitter (BS) 
producing the mapping
\begin{eqnarray}
\nonumber
B^{\dagger}_{I}(\omega_{1})B^{\dagger}_{S}(\omega_{2})
&\mapsto &\frac{1}{2}
[A_1^\dagger(\omega_1) + i A_2^\dagger(\omega_1)][A_2^\dagger(\omega_2) + i A_1^\dagger(\omega_2)]
\end{eqnarray}
whereby $A_{i}^{\dagger}(\omega_{j})$ creates a photon with frequency $\omega_{j}$ in the $i^{th}$ 
exit channel.  
This yields a final Bell state
\begin{eqnarray}
|\bell_{out} \rangle &=& \frac{1}{2} \iint d\omega_{1}  d\omega_{2} {\cal F}(\omega_{1},\omega_{2})
\left(\left(
A_1^\dagger(\omega_1)A_2^\dagger(\omega_2)-A_2^\dagger(\omega_1)A_1^\dagger(\omega_2)\right)\right.\nonumber \\
&+&\left.i\left(A_1^\dagger(\omega_2)A_1^\dagger(\omega_1)+A_2^\dagger(\omega_1)A_2^\dagger(\omega_2)\right)\right)
\Phi(\omega_{1})S(\omega_{2})
|0\rangle.
\end{eqnarray}
The coincidence count rate is determined only by the real part of photon creation term, so we write
\begin{eqnarray}
|\bell_{c} \rangle
&=&
 \frac{1}{2}\int d\omega_{1} \int d\omega_{2}
\left( {\cal F}(\omega_{1},\omega_{2}) \Phi(\omega_{1})S(\omega_{2})\right. \nonumber \\
&-& \left. {\cal F}(\omega_{\textcolor{black}{2}},\omega_{\textcolor{black}{1}}) \Phi(\omega_{\textcolor{black}{2}})S(\omega_{\textcolor{black}{1}})
\right)
|\omega_{1}\omega_{2}\rangle
\end{eqnarray}
and
\begin{eqnarray}
P_{c} &=& \iint d\omega_{1}  d\omega_{2} |\langle\omega_{1}\omega_{2} | \psi_{c} \rangle | ^{2} \nonumber
\\
&=&
\frac{1}{4}\iint d\omega_{1}  d\omega_{2}
\left\{
|{\cal F}(\omega_{1},\omega_{2}) {\cal S}(\omega_{2})|^{2} +
|{\cal F}(\omega_{2},\omega_{1}) {\cal S}(\omega_{1})|^{2}
\right.
\nonumber \\
&-&
\left.
2 {\rm Re}\left[{\cal F}^{*}(\omega_{1},\omega_{2}){\cal F}(\omega_{2},\omega_{\textcolor{black}{1}})
{\cal S}^{*}(\omega_{2}){\cal S}(\omega_{1})
\Phi^{*}(\omega_{1})\textcolor{black}{\Phi}(\omega_{2})
\right]
\right\}.
\label{eq:pc-meth}
\end{eqnarray}
Upon taking $\omega_{1} = \omega_{L} + \omega$ and $\omega_{2} = \omega_{L} - \omega$, 
and changing the integration variable we obtain equation~\ref{eq:pc2} in the text.

\section{Input/Output formalism}
Our theoretical approach is to treat S as a material system interacting with a bath of
quantum photons.  
We shall denote our ``system'' as those degrees of freedom 
describing the material and the photons directly interacting with the sample, 
described by $H_{sys}$
and assume that the photons within sample cavity are exchanged with
external photons in the bi-photon field,
\begin{eqnarray}
H_{r} + H_{rs} = \hbar \int_{-\infty}^{\infty} \left\{
 z B^{\dagger}_{k}(z) B_{k}(z)  
 -i \kappa(z) (\psi_{k}^{\dagger}  B_{k}(z)  -  B_{k}^{\dagger}(z)\psi_{k})\right\} d z,
\end{eqnarray}
where $[B_{k}(z),B_{k'}^{\dagger}(z')] = \delta_{kk'}\delta(z - z')$ are boson operators for 
photons in the laser field, and $[\psi_{k},\psi^{\dagger}_{k}] = \delta_{kk'}$ are boson operators for
cavity photons in the sample that directly interact with the material component of the system.

The Heisenberg equations of motion for the  reservoir and system photon modes are given by 
\begin{eqnarray}
\partial_{t} B_{k}(z) = -i z   B_{k}(z)  +  \kappa(z) \psi_{k}
\end{eqnarray}
and
\begin{eqnarray}
\partial_{t}\psi_{k} =- \frac{i}{\hbar}[\psi_{k},H_{sys}] - \int  \kappa(z) B_{k}(z;t) d z, 
\end{eqnarray}
where the integration range is over all $z$. 
We can integrate formally the equations for the reservoir given either the initial or final states of the reservoir field
\begin{eqnarray}
B_{k}(z;t) = \left\{
\begin{array}{ll}
e^{-iz (t-t_{i})}B_{k}(z;t_{i}) +  \kappa(z)\int_{t_{i}}^{t}ds e^{-iz (t-s)}\psi_{k}(s) & {\rm for\, } t > t_{i} \\
e^{-iz (t-t_{f})}B_{k}(z;t_{f}) -  \kappa(z)\int_{t}^{t_{f}}ds e^{-iz (t-s)}\psi_{k}(s) & {\rm for\, } t < t_{f} .
\end{array}
\right.
\end{eqnarray}
We shall eventually take $t_{i} \to -\infty$ and $t_{f}\to + \infty$ and require that the 
forward-time propagated and reverse-time propagated solutions are the same
at some intermediate time $t$.
If we assume that the coupling is constant over the frequency range of interest, we can 
write 
$$\kappa(z) = \sqrt{\gamma/2\pi}$$
 where $\gamma$ is the rate that energy is exchanged between the reservoir and the system.  This is the (first) Markov approximation.
\end{widetext}

Using these identities, one can find the Heisenberg equations for the 
cavity modes as 
\begin{eqnarray}
\partial_{t}\psi_{k} = -\frac{i}{\hbar}[\psi_{k},H_{sys}] - \sqrt{\frac{\gamma}{2\pi}} \int_{-\infty}^{+\infty}   B_{k}(z;t)dz
\end{eqnarray} 
where $H_{sys}$ is the Hamiltonian for the isolated system.   We can now cast the external field in this equation in terms of its
initial condition:
\begin{eqnarray}
\partial_{t}\psi_{k} &=& -\frac{i}{\hbar}[\psi_{k},H_{sys}] - \sqrt{\frac{\gamma}{2\pi}} \int_{-\infty}^{+\infty}   e^{iz (t-t_{i})}  B_{k0}(z) dz \nonumber \\
&-& {\frac{\gamma}{2\pi}}^{2}\int_{-\infty}^{+\infty}dz \int_{t_{i}}^{t} e^{iz(t-t')}\psi_{k}(t') dt'
\end{eqnarray}

Let us define an input field  in terms of the Fourier transform of the reservoir operators:
\begin{eqnarray} 
 \psi_{k,in}(t) = -\frac{1}{\sqrt{2\pi}}\int_{-\infty}^{+\infty} dz e^{-iz (t-t_{i})}  B_{ko}(z).
\end{eqnarray}
Since these depend upon the initial state of the reservoir, they are essentially a source of stochastic noise for the system.
In our case, we shall use these as a formal means to connect the fields inside the sample to the fields 
in the laser cavity.

For the term involving ${\gamma}/{2\pi}$, 
the integral over frequency gives a delta-function:
\begin{eqnarray}
\int_{-\infty}^{\infty} dz e^{iz(t-t')}  = 2\pi \delta(t-t').
\end{eqnarray}
then 
\begin{eqnarray}
\int_{t_{o}}^{t} dt' \delta(t-t') \psi_{k}(t') = \frac{\psi(t)}{2} {\rm \,\, for\,\,} (t_{o}< t < t_{f})
\end{eqnarray}
This gives the forward equation of motion.
\begin{eqnarray}
\partial_{t} \psi_{k} = -\frac{i}{\hbar}[ \psi_{k},H_{sys}] +\sqrt{\gamma}  \psi_{k,in}(t) - \frac{\gamma}{2}  \psi_{k}(t) 
\end{eqnarray}

We can also define an output field by integrating the reservoir backwards from time $t_{f}$ to time $t$ 
given a final state of the bath, $  B_{kf}$.  
\begin{eqnarray}
  B_{k} = e^{-iz (t-t_{f})}  B_{kf}  - \sqrt{\frac{\gamma}{2\pi}}\int_{t}^{t_{f}}e^{-iz(t-t')}\psi_{k}(t')dt'.
\end{eqnarray}
This produces a similar equation of motion for the output field
\begin{eqnarray}
\partial_{t} \psi_{k} = -\frac{i}{\hbar}[ \psi_{k},H_{sys}] - \sqrt{\gamma }  \psi_{k,out}(t) +  \frac{\gamma}{2}  \psi_{k}(t).
\end{eqnarray}
Upon integration:
\begin{eqnarray}
{\psi_{k,out}(t) = \frac{1}{\sqrt{2\pi}}\int_{-\infty}^{+\infty} d\nu e^{-i\nu (t-t_{f})}  B_{kf}(\nu)}.
\end{eqnarray}

At the time $t$, both equations must be the same, so we can subtract one from the other 
\begin{eqnarray}
  \psi_{k,in} +  \psi_{k,out}  = \sqrt{2\kappa }\psi_{k}
\end{eqnarray}
to produce a relation between the incoming  and outgoing components.   This eliminates the 
non-linearity and explicit reference to the bath modes.  

We now write $\Psi = \{\psi_{k},\psi^{\dagger}_{k}, S_{1},S_{2},\cdots\}$ as a vector of Heisenberg variables for the material system $\{S_{1},S_{2},\cdots \}$ and 
 cavity modes  $\{\psi_{k},\psi^{\dagger}_{k} \}$.
Taking the equations of motion for the all fields to be  linear and of the form
\begin{eqnarray}
\partial_{t} \Psi = {\cal M}_{in}\cdot \Psi  + \sqrt{\gamma} \Psi_{in}
\end{eqnarray}
where  ${\cal M}_{in}$ is a matrix of coefficients which are independent of time.  
The input vector $ \Psi_{in} $ is non-zero for only the terms 
involving the input modes.  We can also write a similar equation in 
terms of the output field; however, we have to account for the 
change in sign of the dissipation terms, so we denote the coefficient matrix as ${\cal M}_{out}$.
In this linearized form,  the forward and reverse equations of motion can be solved formally using the Laplace transform, giving
\begin{eqnarray}
({\cal M}_{in} - iz)  \Psi(z) &=&- \sqrt{\gamma}  \Psi_{in}(z) \\
({\cal M}_{out}-iz)  \Psi(z) &=&+ \sqrt{\gamma}  \Psi_{out}(z).
\end{eqnarray}
These and the relation $  \Psi_{in} +  \Psi_{out}  = \sqrt{\gamma } \Psi $ allows one to eliminate the external variables entirely:
\begin{eqnarray}
 \Psi_{out}(z) = -({\cal M}_{out} - i z )({\cal M}_{in} - iz )^{-1}  \Psi_{in}(z).
 \label{prop}
\end{eqnarray}
This gives a precise connection between the input and output fields. More over, 
the final expression does not depend upon the assumed exchange rate between the
internal $\psi_{k}$ and external $B_{k}(z)$ photon fields.
The procedure is very much akin to the use of M{\o}ller operators in scattering theory. 
To explore this connection, define $\hat\Omega^{(\pm)}_{in} $
as an operator which propagates an incoming solution at $\mp \infty$ to the interaction at time $t =0$
and its reverse  $\hat\Omega^{(\pm)}_{out} $
 which propagates an out-going solution at $\mp \infty$ back to  the interaction at time $t =0$. 
\begin{eqnarray}
\hat \Omega^{(\pm)\dagger}_{in,out}\hat\Omega^{(\pm)}_{in,out} = I,
\end{eqnarray}
and
\begin{eqnarray}
\hat\Omega^{(\pm)}_{in} = ({\cal M}_{in} \mp iz )^{-1} \\
\hat\Omega^{(\pm)}_{out} = ({\cal M}_{out} \pm iz )^{-1}
\end{eqnarray}
Thus, we can write equation~\ref{prop} as 
\begin{eqnarray}
\Psi_{out}(z) = -\hat\Omega^{(-)\dagger}_{out}\hat\Omega^{(+)}_{in}\Psi_{in}(z).
\end{eqnarray}
To compute the response function, we consider fluctuations and excitations from a steady state solution:
\begin{eqnarray}
\Psi(t) = \Psi_{ss}  + \delta\Psi(t).
\end{eqnarray}
The resulting
linearized equations of motion read
\begin{eqnarray}
\frac{d}{dt} \delta \Psi(t) = {\cal M}_{s} \delta \Psi(t)
\end{eqnarray}
implying a formal solution of 
\begin{eqnarray}
\delta \Psi(t) = e^{{\cal M}_{s}t}\delta \Psi(0).
\end{eqnarray}
From this we deduce that the eigenvalues and eigenvectors of ${\cal M}_{s}$ give the fluctuations in terms of the
normal excitations about the stationary solution. 
Using the input/output formalism, we can write the outgoing state (in terms of the Heisenberg variables) in terms of their input values: 
\begin{eqnarray}
 \delta \Psi_{out}(z) = -({\cal M}_{out,s} - i z I )({\cal M}_{in,s} - iz I)^{-1}  \delta\Psi_{in}(z)\nonumber 
\\
\end{eqnarray}
where as given above, $\delta\Psi(z)$ is a vector containing the fluctuations about the stationary values for each of the Heisenberg variables.
The ${\cal M}_{in,s}$ and ${\cal M}_{out,s}$  are the coefficient matrices from the linearisation process.  
The input field satisfies $\langle  \delta \psi_{in}(z)\delta \psi^{\dagger}_{in} (z')\rangle = \delta(z-z')$ and all other terms are zero. 
Thus, the transmission function is given by 
\begin{eqnarray}
\delta(z-z'){\cal S}(z) = \langle \delta \psi^{\dagger}_{out}(z)\delta \psi_{out}(z')\rangle
\label{response}
\end{eqnarray}
In other words, the  ${\cal S}(z)$ is the response of the system to the input field of the incoming 
photon state producing an output field for the out-going photon state.

\section{Dicke model for ensemble of identical emitters}
Let us consider an ensemble of $N$ identical two-level systems corresponding to local molecular sites
 coupled to a set of photon modes described by $\psi_{k}$.
 \begin{eqnarray}
\hat H &=&\sum_{j} \frac{\hbar\omega_{o}}{2}\hat\sigma_{z,j}
+ \sum_{k}\hbar\omega_{k}\hat\psi_{k}^{\dagger}\hat\psi_{k}  \nonumber \\
&+&\sum_{k,j}\frac{\hbar\lambda_{kj}}{\sqrt{N}}(\hat\psi_{k}^{\dagger} + \hat\psi_{k})(\hat\sigma^{+}_{j} + \hat\sigma_j^{-})\label{dickeH}
\end{eqnarray}
where $\{\hat\sigma_{z,j},\hat\sigma^{\pm}_{j}\}$ are local spin-1/2 operators for site $j$, $\hbar\omega_{o}$ is the local 
excitation energy, and $\lambda_{kj}$ is the coupling between the $k$th photon mode and the $j$th site, which we will take to be
uniform over all sites. 
Defining the total angular momentum operators 
$$ \,\, \hat J_{z} = {\sum_j} \hat\sigma_{z,j} \,\,\,{\rm and}\,\,\, \hat J_{\pm} ={\sum_j} \hat\sigma_{j}^{\pm}$$
and
$$\hat J^2=\hat J_z^2+(\hat J_+ \hat J_- + \hat J_- \hat J_+)/2$$ 
as the total angular momentum operator, this Hamiltonian can be cast in the form in equation~\ref{dickeH} by mapping the 
total state space of $N$ spin 1/2 states  onto a 
single angular momentum state vector $|J,M\rangle$.
\begin{eqnarray}
\hat H &=& \hbar\omega_{o}\hat J_{z}+ \sum_{k}\hbar\omega_{k}\hat\psi_{k}^{\dagger}\hat\psi_{k}  \nonumber \\
&+&\sum_{k}\frac{\hbar\lambda_{k}}{\sqrt{N}}(\hat \psi_{k}^{\dagger} + \hat\psi_{k})(\hat J_{+} +  \hat J_{-})
\end{eqnarray}
Note that the ground state of the system corresponds to $|J,-J\rangle$ in which 
each molecule is in its electronic ground state.  Excitations from this state create up to
$N$ excitons within the system corresponding to the state $|J,+J\rangle$.   Intermediate to this
are multi-exciton states which correspond to various coherent superpositions of 
local exciton configurations.   
For each value of the wave vector $k$  one obtains the following Heisenberg  equations of motion for the 
operators
\begin{eqnarray}
\pd{\hat\psi_{k}}{t} &=&     (- i \omega_{k}-\kappa)\hat\psi_{k} - i \frac{\lambda_{k}}{\sqrt{N}}( \hat J_{+} + \hat J_{-}) \\
\pd{\hat\psi_{k}^{\dagger}}{t} &=&  (i \omega_{k}- \kappa)\hat\psi^{\dagger}_{k} + i \frac{\lambda_{k}}{\sqrt{N}}(  \hat J_{+}+ \hat J_{-}) \\
\pd{\hat J_{\pm}}{t}&=& \pm i \omega_{o}\hat J_{\pm} \mp 2i\hat J_{z}\sum_{k}\frac{\lambda_{k}}{\sqrt{N}}( \hat\psi_{k}+\hat\psi_{k}^{\dagger})\\
\pd{\hat J_{z}}{t} &=& +i \frac{\hbar\lambda_{k}}{\sqrt{N}}( \hat J_{-}- \hat J_{+})(\hat \psi_{k}+\hat \psi_{k}^{\dagger}) 
\end{eqnarray}
where $\kappa$ is gives the decay of photon $\hat \psi_k$ into the reservoir.   These are non-linear equations 
and we shall seek stationary solutions and linearize about them. 
 \begin{eqnarray}
 \frac{d}{dt} \delta \Psi(t) = {\cal M}_{s} \delta \Psi(t)
 \end{eqnarray}
 with 
\begin{widetext}
 \begin{eqnarray}
\noindent
  {\cal M}_{s}
 & =& 
  \left[
  \begin{array}{ccccc}
  -(\kappa - i\omega_{k})     &        0                         &    i\lambda_{k}      &  i\lambda_{k}  & 0\\
   0                              & -(\kappa + i\omega_{k})     &    -i\lambda_{k}      &  -i\lambda_{k} & 0\\
  2 i\lambda_{k} \overline{J}_{z}      &          2 i\lambda_{k}\overline{J}_{z}                    &   - i\omega_{o}            &  0                         & 2i\lambda_{k}(\overline{\psi}_{k}+\overline{\psi}_{k}^{\dagger})\\
  -2 i\lambda_{k} \overline{J}_{z}      &            -2 i\lambda_{k} \overline{J}_{z}                   &   0                                &   i\omega_{o}                          & -2i\lambda_{k}(\overline{\psi}_{k}+\overline{\psi}_{k}^{\dagger})\\
     i\lambda_{k}(\overline{J}_{-}-\overline{J}_{+})     &     i\lambda_{k} (\overline{J}_{-}-\overline{J}_{+})        &       i\lambda_{k}(\overline{\psi}_{k}+\overline{\psi}_{k})    & - i\lambda_{k}(\overline{\psi}_{k}+\overline{\psi}_{k})    & 0   \\
   \end{array}
  \right]\label{modelM}
\nonumber \\
 \end{eqnarray}
where $\overline{J}_{z,\pm}$,  $\overline\psi_{k}$, and $\overline\psi_{k}^{\dagger}$ denote the steady state solutions.  
Where we have removed $N$ from the equations of motion by simply rescaling the variables.
The model has both trivial and non-trivial stationary solutions corresponding to the normal and super-radiant regimes. 
For the normal regime,
\begin{eqnarray}
\overline{\psi}_{k} = \overline{\psi}_{k}^{\dagger} = \overline{J}_{\pm,s} = 0
\end{eqnarray}
and 
\begin{eqnarray}
\overline{J}_{z} = \pm \frac{N}{2}.
\end{eqnarray}
which correspond to the case where every spin is excited or in the ground state.  Since we are primarily 
interested in excitations from the electronic ground state, we initially focus our attention to these solutions. 

Non-trivial solutions to these equations predict that above a critical value of the coupling $\lambda > \lambda_{c}$, 
the system will undergo a quantum phase transition to form a super-radiant state. 
It should be pointed out that in the original Dicke model, above the critical coupling, 
the system is no longer gauge invariant leading to a violation of the Thomas-Reiche-Kuhn (TRK) sum rule.
Gauge invariance can be restored; however, the system no longer undergoes a 
quantum phase transition.\cite{Rzazewski:1975}   However, for a {\em driven, non-equilibrium}
system such as presented here, the TRK sum rule does not apply and the quantum phase transition is 
a physical effect. 

\end{widetext}

The non-trivial solutions for the critical regime are given by
\begin{eqnarray}
\overline\psi_{k}^{2} &=& (\overline\psi_{k}^{\dagger})^{2} =  \frac{1}{4}\frac{\omega_{k}\lambda_{k}^{2}}{\omega_{o}\lambda_{c}^{2}}\left(1 - \left(\frac{\lambda_{c}}{\lambda_{k}}\right)^{4}\right)\\
\overline{J}_{\pm}  &=& \frac{1}{2}\left(1 - \left(\frac{\lambda_{c}}{\lambda_{k}}\right)^{4}\right)^{1/2} \\
\overline{J}_{z} &=& -\frac{1}{2}\left(\frac{\lambda_{c}}{\lambda_{k}}\right)^{2}.
\end{eqnarray}

\begin{figure}[t]
\includegraphics[width=\columnwidth]{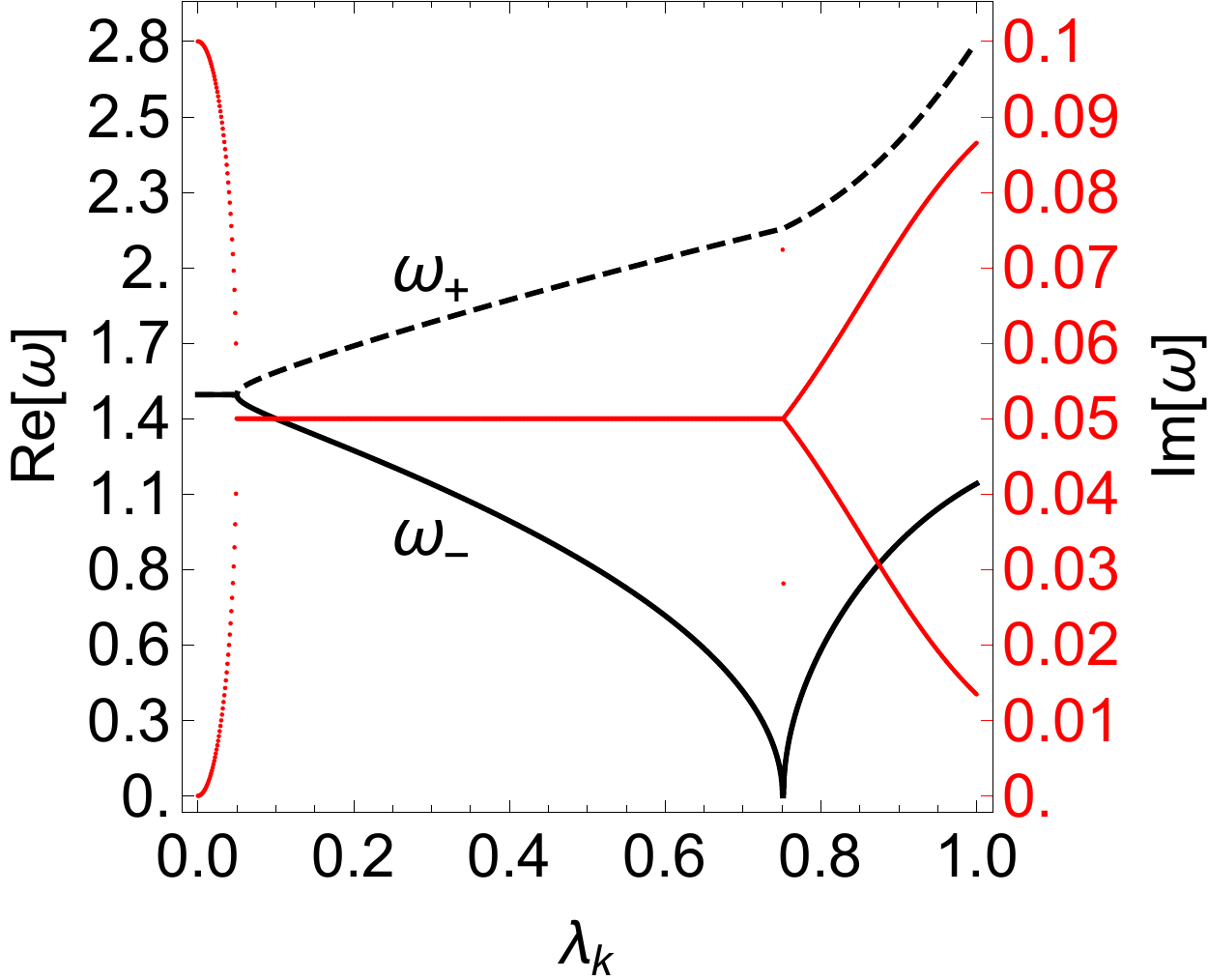}
\caption{
{\bf Upper and lower polariton branches.}  These correspond to the eigenvalues of equation~\ref{modelM} for 
$\omega_{k} = \omega_{o} =1.5 $ and $\kappa = 0.1$ with increasing $\lambda_{k}$ (in scaled units).
The critical coupling occurs at $\lambda_{k} = 0.752$.  
A general feature of this model is that
both photon and exciton-like modes decay at the same rate once $\lambda_{k} = \kappa/2$, corresponding 
to the splitting occurring at $\lambda_{k} = 0.05$. 
}\label{df1}
\end{figure}

The (real) eigenvalues of ${\cal M}_{s}$ gives 4 non-zero and 1 trivial normal mode frequencies (for $\kappa = 0$), 
which we shall denote as 
\begin{eqnarray}
\pm\omega_{\pm} = 
\pm\frac{1}{\sqrt{2}}\sqrt{
(\omega_{k}^{2} + \omega_{o}^{2}) \pm \sqrt{(\omega_{k}^{2}-\omega_{o}^{2})^{2} + 16\lambda_{k}^{2}\omega_{k}\omega_{o}  }
} \nonumber 
\\
\end{eqnarray}
one obtains the critical coupling constant
\begin{eqnarray}
\lambda_{c} = \sqrt{\frac{\omega_{k}\omega_o}{4}\left(1 + \frac{\kappa^{2}}{\omega_{k}^{2}}\right)}
\end{eqnarray}
Figure~\ref{df1} gives the normal mode spectrum for a resonant 
system with $\omega_{k} =  \omega_{o} = 1.5 $ and $\kappa = 0.1$ (in reduced 
units).  

\begin{figure}[ht]
\includegraphics[width=\columnwidth]{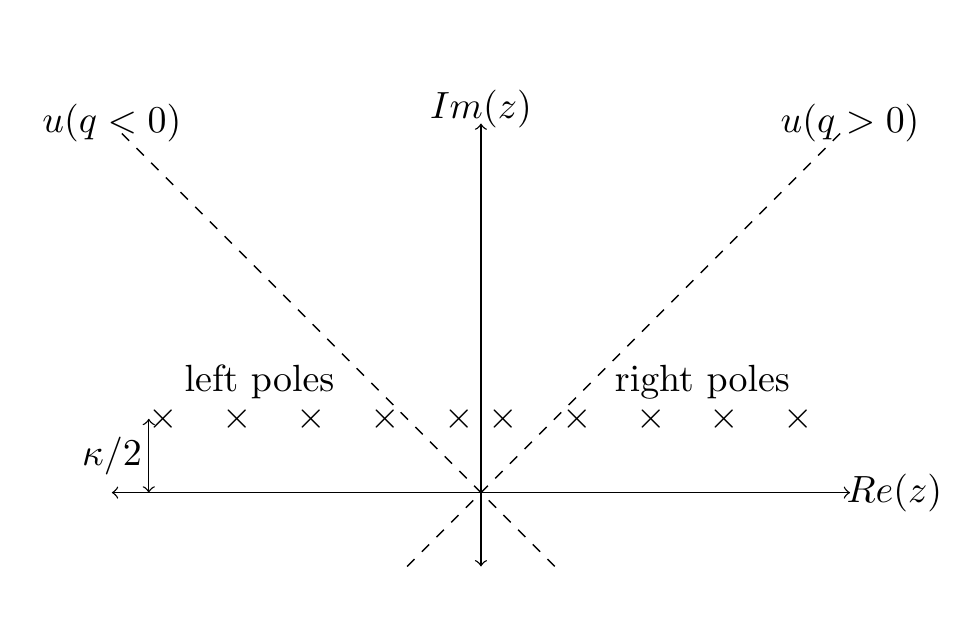}
\caption{{\bf Integration axes and coordinate rotation for integrals \ref{int78}-\ref{int81}.}}\label{complexplane}
\end{figure}

\section{Evaluation of integrals in Eqs. \ref{eq:pc2} and ~\ref{eq:pc}}

The integral in Eq~\ref{eq:pc2} for the photon coincidence can be problematic
to evaluate numerically given the oscillatory nature of the sinc function in ${\cal F}(z)$. 
To accomplish this, we use define a sinc-transformation based upon ${\cal F}(z)$ using the identity
\begin{eqnarray}
{\rm sinc}(z) = \frac{{\rm \sin}(z)}{ z} = \frac{1}{2}\int_{-1}^{1}e^{ikz}dk.
\end{eqnarray}
which yields
\begin{eqnarray}
{\cal F}(z) =\frac{1}{2} \int_{-1}^{1}e^{ik b z^{2}}~dk.
\end{eqnarray}
From this we can re-write each term in Eqs. \ref{eq:pc2} in the form
\begin{eqnarray}
I(t) &=&\int_{-\infty}^{\infty} dz  |{\cal F}(z)|^{2} {\cal S^*}(\omega_L-z){\cal S}(\omega_L+z)e^{2i z t}, \\
&=&
\frac{1}{4} \int_{-1}^{1}dk\int_{-1}^{1} dk'{\cal G}(k-k',t),
\end{eqnarray}
whereby we denote
\begin{eqnarray}
{\cal G}(q,t) = \int_{-\infty}^{\infty} dz~e^{-ibz^{2}q+2itz} {\cal S^*}(\omega_L-z){\cal S}(\omega_L+z).
\label{eq:G}
\end{eqnarray}
The integrand is  highly oscillatory along the $z$-axis; however,
for non-zero $ k-k'= q $, ${\cal G}$  becomes a Gaussian integral under coordinate transformation  obtained by completing the square:
\begin{eqnarray}
-i b q z^2 + 2 i t z &=& - i b q ( z^2 - 2 \frac{t}{b q} z ) \\
 &=& - i b q \left[( z - \frac{t}{b q} )^2 -  (\frac{t}{b q})^2 \right]\\
 &=& - i b q ( z - \frac{t}{b q} )^2 - \frac{t^2}{i b q}.
\end{eqnarray}
For $q>0$, we take $u = (\sqrt{i})( z - \frac{t}{b q} )$, and for $q<0$, we take
$u = (-i\sqrt{i})( z - \frac{t}{b q} )$. 
Solving for $z$ yields:  $z = (-i\sqrt{i}) u +\frac{t}{b q}$ and  $z = (\sqrt{i}) u +\frac{t}{b q}$, respectively. 
In short, the optimal contour of the Gaussian integral is obtained by rotating by $\pi/4$ from the real-axis in the counter-clockwise direction for the
case of $q>0$ and by $\pi/4$ in the clockwise direction for the case of $q<0$ as indicated in Fig.~\ref{complexplane}.

The spectral response ${\cal S^*}(\omega_L-z){\cal S}(\omega_L+z)$ has a number of poles on the complex $z$ plane above the 
real-$z$ axis.  
We now use the residue theorem to evaluate the necessary poles which result as the contour rotates from the real axis to the complex $\pm\pi/4$ axis. 
The 8 second order poles, $\{\rho_{n}\}$, are defined by roots of the denominators  $D(\omega_L-z)$ and $D(\omega_L+z)$
and located $\kappa/2$ above the real axis at  locations symmetrically placed around the origin.
For the counter-clockwise rotation  ($q>0$), 
poles included to the right of the real crossing point will be added and those to the left will be ignored (${\cal P}_{R}(t)$) ; 
whereas for a clockwise rotation ($q<0$), the left-hand poles will be subtracted and the right-hand poles will be ignored (${\cal P}_{L}(t)$).   
\begin{widetext}
For the unique case $q = 0$, all poles are summed (${\cal P}_{all}(t)$).
\begin{eqnarray}
{\cal P}(t) = 2 \pi i \sum_n \lim_{z\to{\rho_{n}}}\frac{d}{dz}\left[(z-\rho_{n})^2
{\cal S^*}(\omega_L-z){\cal S}(\omega_L+z)e^{-i b z^{2} q + 2 i z t }\right].
\end{eqnarray}
  Thus, we obtain  
\begin{eqnarray}
{\cal G}(q>0,t) &=& e^\frac{i t^2}{bq}\int_{-\infty}^{\infty} du e^{-b q u^2} {\cal S^*}(\omega_L-(-i\sqrt{i})u-\frac{t}{b q} ){\cal S}(\omega_L +(-i\sqrt{i})u+\frac{t}{b q} ) \nonumber \\
&+& {\cal P}_{R}(t), \label{int78}\\
{\cal G}(q=0,t>0) &=& {\cal P}_{all}(t), \label{int79} \\
{\cal G}(q=0,t<0) &=& 0,  \label{int80}  \\
{\cal G}(q<0,t) &=& e^\frac{i t^2}{bq}\int_{-\infty}^{\infty} du e^{+b q u^2 } {\cal S^*}(\omega_L-(\sqrt{i})u-\frac{t}{b q} ){\cal S}(\omega_L +(\sqrt{i})u+\frac{t}{b q} )\nonumber \\
&-&  {\cal P}_{L}(t),
\label{int81} 
\end{eqnarray}
whereby ${\cal P}_{L}$, ${\cal P}_{L}$, and ${\cal P}_{all}$ are summed over the left, right, or all poles, respectively.
The resulting expressions are analytic (albeit lengthy) and defined by exponentials and low order polynomials. Completion of the $k$ and $k'$ integrals yield an exact expression for the response. 

\end{widetext}

\bibliographystyle{unsrt}
\bibliography{References-local}

\end{document}